\documentclass[10pt, prd, twocolumn]{revtex4}

\usepackage{amsmath}
\usepackage{amssymb}
\usepackage{graphicx}
\usepackage{bm} 
\usepackage{xcolor}

\begin{document}

\title{
Gravitational field of a pit and maximal mass defects}
\author{Jos\'{e} P. S. Lemos}
\affiliation{Centro de Astrof\'{\i}sica e Gravita\c c\~ao - CENTRA,
Departamento de F\'{\i}sica, Instituto Superior T\'{e}cnico -
IST, Universidade de Lisboa - UL,
Avenida Rovisco Pais 1, Lisboa 1049-001, Portugal, email:
joselemos@ist.utl.pt}
\author{O. B. Zaslavskii}
\affiliation{Department of Physics and Technology, Kharkov
V.~N.~Karazin National University, 4 Svoboda Square, Kharkov 61022,
Ukraine and Institute of Mathematics and Mechanics, Kazan Federal
University, 18 Kremlyovskaya Street, Kazan 420008, Russia,
email: zaslav@ukr.net}

\begin{abstract}

A general relativistic solution, composed of a Zel'dovich-Letelier
interior made of radial strings matched through a spherical thin shell
at radius $r_0$ to an exterior Schwarzschild solution with mass $m$,
is presented. It is the Zel'dovich-Letelier-Schwarzschild star.  When
the radius $r_0$ of the star is shrunk to its own gravitational radius
$2m$, $r_0=2m$, the solutions that appear have very interesting
properties. There are solutions with $m=0$ and $r_0=0$ that further
obey $\frac{2m}{r_0}=1$. These solutions have a horizon, but they
are not exactly black holes, they are quasiblack holes, though
atypical ones. Moreover, the proper mass $m_p$ of the interior is
nonzero and made of one string. Hence, a Minkowski exterior space
hides an interior with matter in a pit. These are the pit solutions.
These pits thus show a maximal mass defect.  There are two classes of
pit solutions, the first encloses a finite string and the second a
semi-infinite one. So these pits are really string pits, which can be
seen as Wheeler bags of gold, albeit totally squashed bags.  There is
also another class, which is a  compact stringy star at the
$\frac{2m}{r_0}=1$ limit with $m$, and thus $r_0$, nonzero.
It is a typical quasiblack
hole and it also shows maximal mass defect. A generic analysis is
presented that shows that pit solutions with $\frac{2m}{r_0}=1$ and
$m=0$ can exist displaying maximal mass defects. The
Zel'dovich-Letelier-Schwarzschild star at the $r_0=2m$ limit is
actually an instance of the generic case.
Notably, these three classes of static solutions yield the same
spectrum of solutions that appear in critical gravitational collapse,
namely, there are solutions that yield naked null singularities, which
here are the two string pit classes of solutions, there are
solutions that yield black holes, which here are represented by the
class of compact stringy stars at the quasiblack hole limit, and the
solutions that disperse away in critical gravitational collapse here
are the static Zel'dovich-Letelier-Schwarzschild stars themselves.  A
thermodynamic treatment of the string pit and stringy star quasiblack
hole solutions is provided, and connections to other solutions are
mentioned.

\end{abstract}



\maketitle

\section{Introduction}

There is the question as to whether, in general relativity, there are
static configurations with some matter interior solution coupled to an
exterior vacuum solution, for which three conditions are satisfied
altogether: The configuration has zero radius $r_0$, $r_0=0$, it also
has zero spacetime, or ADM, mass $m$, $m=0$, and, notwithstanding,
$r_0$ and $m$ obey $\frac{2m}{r_0}$ finite.  We will answer this
question in the positive.  To do so we analyze a generic case, a
generic star composed, say, of a generic interior spherical symmetric
static configuration with a Schwarzschild exterior. Then, as a
nontrivial example, we specifically develop and present the
Zel'dovich-Letelier-Schwarzschild star, which is a solution that
matches the Zel'dovich-Letelier interior made of spherically symmetric
dust strings
up to some junction radius $r_0$
to the exterior Schwarzschild solution. We send $r_0$ to the
gravitational radius of the star $2m$, $r_0\to 2m$, i.e., we take the
highest compact star limit, or the quasiblack hole limit, of the
configuration. Solutions with $m=0$, $r_0=0$, and $\frac{2m}{r_0}=1$,
and thus $\frac{2m}{r_0}$ finite, do indeed appear.  They also have
a nonzero positive proper mass $m_p$.
Thus, clearly these solutions have
maximal mass defects.  Since $m=0$, the exterior spacetime is
Minkowski. But since $m_p$ is not zero, such solutions enclose some
matter in one form or another in a spacetime pit, and are thus pit
solutions.  These pits are atypical quasiblack holes. In the
Zel'dovich-Letelier-Schwarzschild limiting star there are two classes
of pit solutions: The first encloses a finite string and the second a
semi-infinite one. Hence, in these classes the
pits can be described as  string pits.
These string pits resemble Wheeler bags of gold but are totally
squashed.  There is also another class in the
Zel'dovich-Letelier-Schwarzschild limiting star, which is a compact
stringy star at the $\frac{2m}{r_0}=1$ limit with $m$ finite, i.e., a
usual quasiblack hole. These solutions can have a thermodynamic
treatment and have interesting connections to other solutions.

Some comments  are necessary.
(i) The Zel'dovich-Letelier solution was discussed as an interior
general relativistic solution by Zel'dovich \cite{z} in connection
with compact stars, and by Letelier \cite{letelier} as clouds of
strings. It then appeared in the context of global monopoles
\cite{barriolavilenkin} and of string hedgehogs and vacuum bubbles
\cite{guend}. One of the properties of the solutions is that the mass
function $m(r)$ obeys $\frac{2m(r)}{r}=b$, for some fixed $b$ with
$b\leq1$.
Zel'dovich \cite{z} discussed the example of stars
that are about to collapse and so the radial pressure is
irrelevant, whereas Letelier
\cite{letelier} uses energy density $\rho$ and radial pressure $p_r$,
but zero tangential pressure, with equation of state $p_r=-\rho$,
which is also the equation used in \cite{barriolavilenkin,guend}.
This means that one can envisage the matter source as string units
emerging from a common center, and for this reason the source is
called string dust, with the Zel'dovich setup being a particular
situation of string dust, i.e., pure dust, since $p_r=0$. In brief,
string dust has three features, namely, radial strings, $p_r=-\rho$,
and no tangential pressure. We are interested in string dust solutions
as interior solutions.  We make use of the junction formalism
\cite{isr} to match a Zel'dovich-Letelier interior to a Schwarzschild
exterior and obtain the Zel'dovich-Letelier-Schwarzschild star.
(ii) Objects for which the surface radius $r_0$ of the matter, e.g., a
star, is at its own gravitational radius, $r_0=2m$, are
the highest compact
stars called quasiblack holes \cite{qbh1}.  Quasiblack holes are on
the verge of becoming black holes, have special properties, and are
relatives to both black holes and null naked singularities
\cite{qbh1,qbh2,qbh3,qbh4,qbh5}.
(iii) The answer to the question as to whether, in general relativity,
zero mass $m$ at a point $r=0$ can nonetheless have a quotient
$\frac{2m}{r}$ finite is known to be yes in a dynamical
setting. Indeed, the inhomogeneous dust spherical collapse of the
Lema\^itre-Tolman-Bondi models can produce an $m=0$, $r=0$, naked null
singularity.  At the critical moment the density has a $\frac{1}{r^2}$
isothermal profile, and $\frac{2m(r)}{r}$ is finite, indeed
$\frac{2m(r)}{r}=1$ at the center \cite{lemos1,lemos2,lemos3}.  Also,
Choptuik collapse of a scalar field \cite{choptuik} allows for a
critical case, which divides expansion of the scalar field back to
infinity from collapse of the scalar field to a black hole, where a
zero mass and zero radius black hole, with $\frac{2m}{r}=1$, and thus
$\frac{2m}{r}$ finite, forms. This can also be interpreted as the
formation of a naked null zero mass singularity.
(iv) Mass defects with maximal values have appeared in specific
nonstationary models \cite{ruban} where one can have matter with an
infinite amount of interior proper mass, but the exterior spacetime
has zero spacetime mass \cite{mass}.  This is the maximal possible
gravitational mass defect.
(v) Bag of gold solutions appeared in \cite{wheeler}, see also
\cite{ong}. The Wheeler bags of gold are exemplified by a closed FLRW
universe glued to the other side of a Schwarzschild black hole through
an Einstein-Rosen bridge with a knot at the junction closing the bag.
(vi) Since the string pit and stringy star compact solutions found are
quasiblack holes, it is of interest to discuss their thermodynamics as
in \cite{ent,ext}.
(vii) There are several related interesting solutions to the
Zel'dovich-Letelier interior that have the mass function $m(r)$ obeying
$\frac{2m(r)}{r}=b$ but are not string dust, i.e., the equation of
state differs from $p_r=-\rho$, see,
e.g.,~\cite{klein,miszapo,chandra,chavanis,bg,dymnikova1992,nl,
berezin}.
A match to a vacuum exterior of these related solutions does not yield
Zel'dovich-Letelier-Schwarzschild stars.

The paper is organized as follows.
In Sec.~\ref{basic} we lay down the spacetime's basic features and
arrive naturally at the concept of spacetime pits, i.e., solutions
with $m=0$ and $\frac{2m}{r_0}$ finite which yield maximal mass
defects.
In Sec.~\ref{zls} we display the Zel'dovich-Letelier solution and
make a proper matching to find the Zel'dovich-Letelier-Schwarzschild
star.
In Sec.~\ref{lim} we take the quasiblack hole limit $r_0\to 2m$ of the
Zel'dovich-Letelier-Schwarzschild star and find three classes  of  objects
with the highest compactification.  The first two classes are pit solutions, more
precisely, string pit solutions, one enclosing a finite string, and the
other a semi-infinite string, both with $m=0$, $\frac{2m}{r_0}=1$, and
maximal mass defects.  The third class is a  compact stringy
star, more
precisely, a string star with highest compactness, i.e.,
$\frac{2m}{r_0}=1$, finite $m$, and also a maximal mass defect.
In
Sec.~\ref{c} we conclude, giving a synopsis
with the results in a table, glimpsing through the
thermodynamics of the string pit and stringy star solutions,
and connecting with related work by others.


\section{Basic features of spacetimes with a pit and maximal mass
defect}
\label{basic}

\subsection{Spacetime generics}

A general static spherical symmetric
spacetime with spacetime coordinates $(t,r,\theta,\phi)$
has a line element that can be written
in the form
$
ds^2=-\left(1-\frac{2m(r)}{r}\right){\rm e}^{2\psi(r)}dt^2
+\frac{dr^2}{1-\frac{2m(r)}{r}}+r^2d\Omega^2\,,
$
where $m(r)$ and $\psi(r)$ are functions of $r$,
and $d\Omega^2$ is the line element on
the unit sphere, $d\Omega^2=d\theta^2+\sin^2\theta\,d\phi^2$,
and $\theta$ and $\phi$ are the angles on it.
Assume that for $r\leq r_0$, for some radius $r_0$,
there is a fluid with
energy-momentum tensor $T^{ab}$ given by
${T^{a}}_b={\rm diag}(-\rho,p_r,p_t,p_t)$,
where $\rho$ is the fluid's energy density, $p_r$
its radial pressure, and $p_t$ its tangential pressure,
all functions of $r$. Then the Einstein equation
of general relativity
$G_{ab}=8\pi T_{ab}$, where 
$G_{ab}$ is the Einstein tensor, and we put
the constant of gravitation and the velocity of light to unity,
yields
$m(r)=4\pi \int_{0}^{r}dr^\prime {r^\prime}^2\rho(r^{\prime})$
and $\psi(r) =4\pi \int_{r_0}^{r}dr^{\prime}\,r^{\prime}\,
\frac{\rho(r^{\prime})+p_r(r^{\prime})}
{1-\frac{2m(r\prime)}{r\prime}}$.
There is yet another equation involving the 
tangential pressure $p_t$ that we do not need
right now. In the model that we are going to use
for the interior one has $\rho(r)+p_r(r)=0$
so that $\psi(r)=0$ throughout.
Thus, in this case the line element that we start with
in the $(t,r,\theta,\phi)$
coordinates reduces to 
\begin{equation}
ds^2=-\left(1-\frac{2m(r)}{r}\right)dt^2
+\frac{dr^2}{1-\frac{2m(r)}{r}}+r^2d\Omega^2\,,
\label{metric2}
\end{equation}
where 
\begin{equation}
m(r)=4\pi \int_{0}^{r}dr^\prime {r^\prime}^2\rho(r^{\prime}) \,,
\label{massenergyr}
\end{equation}
is now the only metric
function, usually
called the
mass function
and defined  for $r\leq r_0$.

Other functions of interest here are 
the proper mass $m_p(r)$, the
proper distance $l_p(r)$ 
from the center to any $r\leq r_0$,
the area $A(r)$
of a constant $r$ sphere,
and the proper volume $V_p(r)$.
The proper mass $m_p(r)$ is defined as
\begin{equation}
m_p(r)=4\pi \int_{0}^{r}dr^\prime
\frac{{r^\prime}^2\rho(r^\prime)}{\sqrt{1-
\frac{2m(r^\prime)}{r^\prime}}}\,,
\label{massproperr}
\end{equation}
the proper distance $l_p(r)$ from the center to any $r$ is
defined as
\begin{equation}
l_p(r)=\int_{0}^{r}
\frac{dr^\prime}{\sqrt{1-\frac{2m(r^\prime)}{r^\prime}}}\,,
\label{lpr}
\end{equation}
the area $A(r)$ of a constant $r$ sphere is 
\begin{equation}
A(r)=4\pi r^2\,,
\label{Ar}
\end{equation}
and the proper volume is defined by
\begin{equation}
V_p(r)=4\pi \int_{0}^{r}dr^\prime
\frac{{r^\prime}^2}{\sqrt{1-\frac{2m(r^\prime)}{r^\prime}}}\,.
\label{Vr}
\end{equation}
These functions at the boundary $r_0$ become specific
important quantities and we put
\begin{eqnarray}
&&m\equiv m(r_0)\,,\quad
m_p\equiv m_p(r_0)\,,\nonumber\\
&&l_p\equiv l_p(r_0)\,,\quad
A\equiv A(r_0)\,,\quad\;
V_p\equiv V_p(r_0)\,.
\label{r0quantities}
\end{eqnarray}

At $r_0$ there is a boundary that can be smooth or can have a shell.  If
there is a shell it can have zero or nonzero proper mass and zero or
nonzero pressure.

For $r\geq r_0$ we assume that the solution is vacuum and thus that it
is the Schwarzschild solution, $m(r)=M$ constant, where $M$ is the
spacetime mass or energy.  In general $M$ and
$m$ are different.  Here we work with the case $M=m$ as we will see it
is the case in the matching of the Zel'dovich-Letelier interior
solution to the exterior Schwarzschild solution. Thus, $m$ is the 
mass of the spacetime.

There are two characteristic masses in this setting,
the spacetime mass $m$ and the proper mass of the
object $m_p$.
It is then appropriate to define
generically the mass defect $\Delta m$ of an
object as
\begin{equation}
\Delta m=m_p-m\,, 
\label{massdefect}
\end{equation}
which indicates how much mass, or energy, was put
into the construction of the spacetime.

\subsection{Features of pit spacetimes with
$m=0$, $r_0=0$, $\frac{2m}{r_0}=1$, and maximal mass defect}

We put $M=m$, i.e., there is no contribution
to the exterior spacetime mass from the
boundary at $r_0$, and stick to calling it $m$.
Considering the mass function $m(r)$
appearing in Eq.~(\ref{metric2}),
we assume that 
$1-\frac{2m(r)}{r}$ is uniformly bounded and write
$1-\frac{2m(r)}{r}\geq0$, i.e.,
\begin{equation}
\frac{2m(r)}{r}\leq1\,.
\label{ineq1}
\end{equation}
With this assumption we can
make some general, concrete remarks.
Define $\varepsilon$ as any positive
number, which  can be as small as we want,
and $\chi(r)$ a function
of $r$ always greater than zero,
such that $1-\frac{2m(r)}{r}=\varepsilon \chi(r)$.
Take the maximal value of $\chi(r)$
as $\chi_{\max}$ and its minimum value as
$\chi_{\min}$. 
Then, since 
Eq.~(\ref{ineq1}) holds, the integral of
Eq.~(\ref{massenergyr})
converges, 
and taking the integrals up to
the boundary $r_0$
in
Eqs.~(\ref{massenergyr}) and 
(\ref{massproperr})
leads to
\begin{equation}
\frac{m}{\sqrt{\varepsilon\chi_{\max} }}
\leq m_p\leq
\frac{m}{\sqrt{\varepsilon\chi_{\min} }}\,.
\label{twoineq}
\end{equation}
Take the quasiblack hole limit, i.e., $r_0\to2m$, or
$\frac{2m}{r_0}\to1$ from below,
so that one also has
$\varepsilon\to0$.  This is a configuration made of some
material with boundary radius at its own gravitational radius $2m$, it
is a configuration on the verge of becoming a black hole.  Suppose
that $m_p$ remains finite on this limit.  Then, since
$\varepsilon\to0$, one has mandatorily
from Eq.~(\ref{twoineq}) that $m$ goes to zero, so, since $r_0=2m$
in this limit, one also has $r_0\rightarrow 0$.  Thus, one has an 
object that has $m=0$, i.e.,
zero mass energy $m$, and
$r_0=0$, i.e.,  zero radius $r_0$, with
$\frac{2m}{r_0}=1$, and also has finite nonzero proper mass $m_p$.  In
addition, defining
the mass defect $\Delta m=m_p-m$ as in Eq.~(\ref{massdefect}),
we see that this object has a maximal mass defect given by
$\Delta m=m_p$.  In brief, such an object has
\begin{equation}
m=0\,,\quad
r_0=0\,,\quad \frac{2m}{r_0}=1\,,\quad
m_p={\rm finite}\,,\quad \Delta m=m_p\,.
\end{equation}
This is an amazing object.  It has zero spacetime
mass and zero area radius,
and, although $m=0$, the ratio $\frac{2m}{r_0}$ is not zero, but
actually one.  In addition, it has finite proper mass and maximal mass
defect.  We call this structure a pit, as it stores a nonzero proper
mass in a zero spacetime mass spacetime with zero area radius.  As
$\frac{2m}{r_0}=1$, the pit is indeed a quasiblack hole, although an
atypical one. A specific realization of this general analysis is
through the Zel'dovich-Letelier-Schwarzschild star that we will
display next and where there are three possible classes, two of them
being string pits, each  with distinct and rather interesting
features, and the other being a compact stringy star at the quasiblack
hole state.

\section{The Zel'dovich-Letelier-Schwarzschild star
and its limits}
\label{zls}

\subsection{The interior, the shell junction, the exterior, and the
Zel'dovich-Letelier-Schwarzschild star}
\label{iesf}

\subsubsection{The Zel'dovich-Letelier interior}
Let us be concrete.
To simplify, let us choose an equation of state of the form
$
p_r=-\rho
$.
Then, inside for $r\leq r_0$,
we have  that indeed $\psi(r)=0$ and
the only function that matters
is the function
$m(r)$ that appears in Eqs.~(\ref{metric2}) and~(\ref{massenergyr}).
The conservation law ${T^{ab}}_{;b}=0$ with $a=r$ gives
$p_{t}=p_r+\frac{r}2p_r^{\prime}$,
and using the equation of state
$
p_r=-\rho
$
one gets $p_{t}=-\rho -\frac{r}2\rho^{\prime}$.
Following Zel'dovich \cite{z}
and Letelier \cite{letelier}, see also 
\cite{barriolavilenkin,guend}, we put
$\rho =\frac{b}{8\pi r^2}$,
where $b$ is a positive constant. Thus, the full
general relativistic solution
using the Einstein equation is
\begin{equation}
\rho =\frac{b}{8\pi r^2}\text{,}  \label{rho}
\end{equation}
\begin{equation}
p_r =-\frac{b}{8\pi r^2}\text{,}  \label{pr}
\end{equation}
\begin{equation}
p_t =0\text{.}  \label{pt}
\end{equation}
Since $p_t=0$, the source is string dust, strings in
the radial direction
up to $r_0$.
Putting the energy-density expression Eq.~(\ref{rho})
into Eq.~(\ref{massenergyr}),
one obtains $m(r)=\frac{b}2r$, i.e., $\frac{2m(r)}{r}=b$,
and the line element,
Eq.~(\ref{metric2}), becomes
\begin{equation}
ds^2=-\left(1-b\right)dt^2+\frac{dr^2}{1-b}+r^2d\Omega ^2\text{.}
\label{metricusu}
\end{equation}
This metric yields a spacetime that has a spherical
conic deficit. Indeed, redefining ${\bar t}=\sqrt{1-b}\,t$
and ${\bar r}=\frac{r}{\sqrt{1-b}}$, one gets
the conical form of the metric, namely,
$
ds^2=-d{\bar t}^2+d{\bar r}^2+{\bar r}^2\left(1-b\right)
d\Omega ^2\text{}
$.
Clearly, the inside metric is a deficit angle metric.  Returning to
the main functions of a static spherical symmetric spacetime, 
Eqs.~(\ref{massenergyr})-(\ref{Vr}), we can put them in the case of
the Zel'dovich-Letelier spacetime in the form
\begin{equation}
m(r)=\frac12 {b}r\,,  \label{mrzl}
\end{equation}
\begin{equation}
m_p(r)=
\frac{br}{2\sqrt{1-b}}\,,
\label{mprzl}
\end{equation}
\begin{equation}
l_p(r)=\frac{r}{\sqrt{1-b}}\,,
\label{lprzl}
\end{equation}
\begin{equation}
A(r)=4\pi r^2\,,
\label{Ar2}
\end{equation}
\begin{equation}
V_p(r)=\frac{4\pi r^{3}}{3\sqrt{1-b}
}\,.
\label{Vrzl}
\end{equation}
Clearly, we can rewrite $m_p(r)$ as
$m_p(r)
=\frac{m(r)}{\sqrt{1-b}}$,
$l_p(r)$ as
$l_p(r)=\frac{2m(r)}{b\sqrt{1-b}}=\frac{2m_p(r)}{b}$,
and $V_p(r)$ as
$V_p(r)=\frac{32m(r)^{3}}{3b^{3}\sqrt{1-b}}=
\frac{32m_p(r)^{3}(1-b)}{3b^{3}}$.
It is assumed
that $b\leq1$ so that
the metric in Eq.~(\ref{metricusu}) is static,
and in addition it is assumed that the parameter $b$
is positive, so that
the mass function $m(r)$ in Eq.~(\ref{mrzl})
is positive, 
i.e., we put
\begin{equation}
0<b\leq1\,.
\label{bcond}
\end{equation}
Equations.~(\ref{metricusu})-(\ref{Vrzl})
with the condition (\ref{bcond})
characterize the interior spacetime defined for
$r\leq r_0$.

Note that the interior solution is singular at $r=0$, the density and
radial pressure, given in Eqs.~(\ref{rho}) and~(\ref{pr}),
respectively, diverge there, and thus the Ricci and Riemann tensors and
corresponding scalars diverge. Zel'dovich \cite{z} deals with
gravitational collapse issues, neglects $p_r$, and dismisses this
singularity problem, showing that rounding up the energy density $\rho$
at the origin makes no difference for his final results.  Letelier
\cite{letelier} suggests that the solution can be used as an
intermediary solution between the Schwarzschild interior solution and
a Schwarzschild exterior.  Here we use the solution to match it to a
Schwarzschild exterior, and in taking the limit $r_0\to2m$ it is found
that this singularity is not naked because it is within a quasiblack
hole.

\subsubsection{The shell junction}

The junction of the inside and the outside is at some $r_0$.  At the
junction $r_0$ we consider the metric to be of the form
\begin{equation}
ds^2=-d\tau^2+r_0^2\,d\Omega ^2\text{,}
\label{metricjunction}
\end{equation} where $\tau$
is the proper time
at the junction.  For the outside we consider a vacuum spacetime and
thus from Birkhoff's theorem it is the Schwarzschild spacetime.  Since
from the inside the radial pressure at $r_0$ is nonzero, namely,
$p_r=-\frac{b}{4\pi r_0^2}$, see Eq.~(\ref{pr}), and from the outside
$p_r=0$, as we consider that the outside is vacuum, there is a clear
jump in the radial pressure that has to be smoothed out by a thin
spherical shell at the junction at $r_0$.  The energy-momentum tensor
of the thin shell can be found.  We write the contribution to the
energy-momentum tensor
$
{T^{a}}_{b}
$
from the
shell in the form
$
{T^{a}}_{b}={S^{a}}_{b}\,\delta (l-l_{0})
$,
where ${S^{a}}_{b}$
is the intrinsic energy-momentum tensor associated with the shell,
$\delta$ is the Dirac delta function,
$l$ is the proper radial length in the
neighborhood of the shell, and $l_{0}$ corresponds to the
boundary at the shell.
Then, following the junction formalism
for general relativity \cite{isr},
one finds ${S^{\tau}}_{\tau}=0$ and
${S^{\theta}}_{\theta}={S^{\phi}}_{\phi}=
\frac{b}{16\pi r_0\sqrt{1-b}}$.
Assuming that the shell is made of a perfect fluid
and
writing ${S^{\tau}}_{\tau}\equiv-\sigma$
and ${S^{\theta}}_{\theta}={S^{\phi}}_{\phi}\equiv P$, where
$\sigma$ is the energy density of the shell and
$P$ is the tangential pressure at the shell,
we thus have  
\begin{equation}
\sigma=0\,,\quad\quad P=\frac{b}{16\pi
r_0\sqrt{1-b}}\,.
\label{shellstresses}
\end{equation}
Note that there is no  mass for the shell
as $m_{\rm shell}=4\pi r_0^2\,\sigma=0$. Note that $P$ 
closes the conical deficit set in by the interior
spacetime 
such that the exterior Schwarzschild
spacetime has no conical deficit. The shell's tangential
pressure
$P$ is there to close ends, literally.

\subsubsection{The exterior Schwarzschild}

The outer spacetime is vacuum, and therefore
the exterior general relativistic
metric, the metric for $r\geq r_0$,
is Schwarzschild, i.e.,
$ds^2=-\left(1-\frac{2M}{r}\right)dt^2+
\frac{dr^2}{1-\frac{2M}{r}}+r^2d\Omega ^2$
for some spacetime mass
$M$. In general $M$ and  $m$
have different values. Here,
since the shell 
has no mass,
$m_{\rm shell}=0$,
we  deal with the case $M=m$ and keep $m$
throughout. Thus, we put
\begin{equation}
ds^2=-\left(1-\frac{2m}{r}\right)dt^2+
\frac{dr^2}{1-\frac{2m}{r}}+r^2d\Omega ^2\,,
\label{metricschw}
\end{equation}
as
 the
exterior Schwarzschild metric.

\subsubsection{The full solution:
The Zel'dovich-Letelier-Schwarzschild star}

The full general relativistic
solution is composed of three parts. The inside
with the metric given by Eq.~(\ref{metricusu}),
the
shell with the metric given by Eq.~(\ref{metricjunction}), and 
the outside
with the metric given in Eq.~(\ref{metricschw}).
The main global features can be found at the
junction $r_0$.
From Eqs.~(\ref{mrzl})-(\ref{Vrzl})
they are
\begin{equation}
m=\frac12 b\,r_0\,,
\label{massattheboundary0}
\end{equation}
\begin{equation}
m_p=\frac{br_0}{2\sqrt{1-b}}\,,
\label{mp0}
\end{equation}
\begin{equation}
l_p=\frac{r_0}{\sqrt{1-b}}\,,
\label{lp0}
\end{equation}
\begin{equation}
A=4\pi r_0^2\,,
\label{area0}
\end{equation}
\begin{equation}
V_p=\frac{4\pi r_0^{3}}{3\sqrt{1-b}
}\,.
\label{V0}
\end{equation}
We can rewrite $m_p$ as
$m_p
=\frac{m}{\sqrt{1-b}}$,
$l_p$ as
$l_p=\frac{2m}{b\sqrt{1-b}}=\frac{2m_p}{b}$,
and $V_p(r)$ as
$V_p=\frac{32m^{3}}{3b^{3}\sqrt{1-b}}=
\frac{32m_p^{3}(1-b)}{3b^{3}}$, and
we recall that Eq.~(\ref{bcond}) should be taken
into account,
i.e., $0<b\leq1$.
This is the full Zel'dovich-Letelier-Schwarzschild spacetime solution,
i.e., the Zel'dovich-Letelier-Schwarzschild star.

\subsection{The limit $r_0\rightarrow 2m$
of the Zel'dovich-Letelier-Schwarzschild star:
Distinguishing features}

We are interested in the limit in which
\begin{equation}
r_0\rightarrow 2m\,, \label{lr02m}
\end{equation}
i.e., the
quasiblack hole limit in which an object is at its own gravitational
radius
\cite{qbh1,qbh2,qbh3,qbh4,qbh5}.
It follows from Eq.~(\ref{massattheboundary0}) that this
means
\begin{equation}
b\rightarrow 1\,. \label{bto1}
\end{equation}

Let us take $m_p$ in Eq.~(\ref{mp0}) as the
quantity that identifies the possible different classes.
To see this, we put $m_p$ as $m_p(1-b)^{\gamma}=\mu$ for some
exponent $\gamma$
and some finite
renormalized  proper mass $\mu$, with $\mu\geq0$.
This choice for the relation between
the proper mass $m_p$ and the renormalized
mass $\mu$ is taken because it takes care of
all the independent cases, actually three cases,
when one takes the limit given in Eq.~(\ref{bto1}).
For now we leave $b$ generic, only
afterward do we take that limit.
Then Eqs.~(\ref{massattheboundary0})-(\ref{V0})
with $r_0=2m$ of Eq.~(\ref{lr02m})
give
\begin{equation}
m=\mu (1-b)^{\frac{1}{2}-\gamma }\text{,}
\label{massattheboundary}
\end{equation}
\begin{equation}
m_{p}=\mu (1-b)^{-\gamma }\text{,}
\label{mp}
\end{equation}
\begin{equation}
l_p=2\mu\,\frac{(1-b)^{-\gamma }}{b}\,,
\label{lp}
\end{equation}
\begin{equation}
A=16\pi\mu^2\, \frac{(1-b)^{1-2\gamma}}{b^2} \,,
\label{area}
\end{equation}
\begin{equation}
V_{p}=\frac{32}{3}\mu ^{3}\,\frac{(1-b)^{1-3\gamma}}{b^{3}}\,,
\label{V}
\end{equation}
respectively.
From Eqs.~(\ref{massattheboundary}) and~(\ref{mp})  
we see that a negative exponent $\gamma$ gives zero mass
$m$ and zero proper mass $m_p$
in the limit of Eq.~(\ref{bto1}). It is thus of no interest
as it gives nothing, and we impose $\gamma\geq0$. 
From Eq.~(\ref{massattheboundary}) we see
that an exponent $\gamma$ greater that $\frac12$ gives an infinite
mass $m$ in the limit of Eq.~(\ref{bto1})
and the spacetime with the line element
given in Eq.~(\ref{metricschw}) is not well defined, 
and we impose $\gamma\leq\frac12$.
Thus, $\gamma$ is within the range
\begin{equation}
0\leq\gamma\leq\frac12\,.
\label{range}
\end{equation}
Then Eqs.~(\ref{massattheboundary}) and~(\ref{mp})
show that there are three distinct main classes:
A.
$\gamma=0$, which yields $m=0$ and
$m_p$ equal to $\mu$, and thus $m_p$ is finite.
It is a string pit solution;
B. $0<\gamma<\frac12$, which yields $m=0$
and $m_p$ infinite. It is also a string pit solution
with different properties;
C. $\gamma=\frac12$, which yields $m$ finite and $m_p$ infinite. It is
a compact stringy star with the highest compactification, it
is not a pit.
Let us analyze these three
classes in detail.


\section{The three limiting solutions:
Two string pits and a  compact stringy star}
\label{lim}

\subsection{A finite string in a pit, i.e., a string pit,
almost detached from spacetime
hanging from a point}

Here we find a finite string in a pit, a string pit,
almost detached from spacetime
hanging from a point, indeed with the spacetime mass $m=0$ and the
proper mass $m_p={\rm finite}$. This is the class $\gamma=0$.

We are interested in the limit in which $r_0\rightarrow 2m$, see
Eq.~(\ref{lr02m}), i.e., the quasiblack hole state.  It follows from
Eq.~(\ref{massattheboundary0}) that this implies $b\rightarrow 1$, see
Eq.~(\ref{bto1}).  We put $\gamma=0$ in
Eqs.~(\ref{massattheboundary})-(\ref{V}) and analyze the spacetime's
main features.  From Eq.~(\ref{massattheboundary}) we have
$m\rightarrow 0$, i.e., $m=0$ in the limit.  Equation~(\ref{mp})
yields $m_p=\mu$, so $m_p$ is finite, with the subcase $m_p=0$ being a
trivial case.  From Eq.~(\ref{lp}), the total proper length $l_p$
remains finite. This is clear as $l_p=2m_p$ and $m_p$ is finite, see
Eq.~(\ref{lp0}) with $b=1$.  From Eq.~(\ref{area}), the surface area
is $A=0$, and the area radius of the boundary is $r_0=0$.  From
Eq.~(\ref{V}), the proper volume is zero, $V_p=0$.

\begin{figure}[h]
\begin{center}
\includegraphics[scale=0.5]{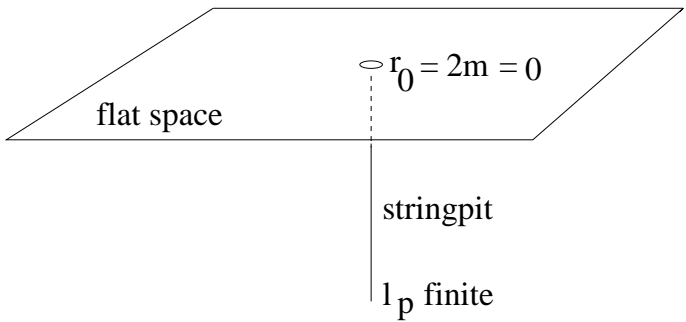}
\caption{
A $t={\rm constant}$ and $\theta=\frac\pi2$ space representation of
the spacetime given by the Zel'dovich-Letelier-Schwarzschild star with
proper mass $m_p=\mu={\rm finite}$; actually $m_p(1-b)^{\gamma}=\mu$,
with $\gamma=0$ and $b=1$ and spacetime mass $m=0$.  This class has
$r_0=2m$, and so it is a quasiblack hole, an atypical one, as it
satisfies $r_0=2m=0$.  The space inside is a region of matter packed
at the highest level, composed of a pit made of a one-dimensional
string with finite proper length, hung from a point with $r_0=0$,
which opens up to a massless $m=0$ Minkowski spacetime, i.e., a flat
space here.  The point $r_0=0$ yields the singular horizon of the
quasiblack hole and joins the almost detached string to the rest of
the space.  Note that although $m=0$ and $r_0=0$ their ratio is finite
as $\frac{2m}{r_0}=1$.  This object has maximal mass defect.  The
representation of this class of string pit solution shows clearly that
the solution is a totally squashed Wheeler bag of gold.
}
\label{case1}
\end{center}
\end{figure}

Thus, the full spacetime can be understood as follows.  The inside
solution is made of a one-dimensional open string, with finite length
and zero volume.  That the inside spacetime is a one-dimensional
string can also be seen from the conical form of the inside metric,
where for $b=1$ and $\bar r$ finite, as  is the case, the angular
part disappears leaving a one-dimensional space, i.e., a
two-dimensional spacetime.  This single string in the inside
spacetime pit
is what is left from the hedgehog continuous spherical distribution of
strings in the original Zel'dovich-Letelier interior solution.  It is a
finite string almost detached from spacetime hanging from a point.
For the shell, which joins the inside and the outside, one deduces it
is now a point as $r_0=0$.  Then, from Eq.~(\ref{shellstresses}),
since $b=1$ and $r_0=0$, the tangential stresses tend to infinity,
$P\rightarrow \infty$, and thus the point $r_0=0$ is singular, a type
of singular horizon.
For the outside, one has that the spacetime is Minkowski as $m=0$.
Thus, in a nutshell, a Minkowski exterior spacetime
hides a finite string pit. 
For a $t={\rm constant}$ and, e.g., $\theta=\frac\pi2$ space
representation of the spacetime, see Fig.~\ref{case1}, where it is
clear that the packed region with matter is a pit with a string
hanging in the middle of flat space.

Note five important and interesting properties of this class of
string pit solution.  First, the interior mass of the
Zel'dovich-Letelier-Schwarzschild star
\cite{z,letelier,barriolavilenkin,guend,isr} in this limit is hidden
to the outside as it does not manifest itself gravitationally to the
outer space since $m=0$.  Second, it is also hidden because it is
invisible since it is a quasiblack hole
\cite{qbh1,qbh2,qbh3,qbh4,qbh5}. It is thus invisible for two
reasons.  Third, although $m=0$, its ratio to $r_0$ is finite, indeed,
$\frac{2m}{r_0}=1$. These three features characterize an atypical
quasiblack hole. Thus, the dynamical gravitational collapse setting in
\cite{lemos1,lemos2,lemos3,choptuik} for which a null naked
singularity, i.e., a singular horizon, forms when $m=0$ at $r=0$ and
$\frac{2m}{r}=1$ is also established in the static case that we are
analyzing.  Fourth, the mass defect, i.e., the proper mass minus the
energy of the assembled object given
in Eq.~(\ref{massdefect}) is $\Delta m=m_p-m=m_p$,
so we are in the presence of an object
with maximal mass defect, see also \cite{ruban,mass}.  Fifth,
it is a Wheeler bag of gold \cite{wheeler,ong} but is totally squashed.

For a study of the geodesics in this spacetime, see the 
Appendix~\ref{geo}.


\subsection{A semi-infinite string in a pit, i.e., a string pit,
almost detached from spacetime hanging from a point}

Here we find a semi-infinite string in a pit, a semi-infinite
string pit, almost detached from spacetime hanging from a point, with
indeed the spacetime mass $m=0$ and the proper mass $m_p=\infty$. This
is the class $0<\gamma<\frac12$.

We are again interested in the limit in which $r_0\rightarrow 2m$, see
Eq.~(\ref{lr02m}), i.e., the quasiblack hole state.  It follows from
Eq.~(\ref{massattheboundary0}) that again this implies that $b\rightarrow
1$, see Eq.~(\ref{bto1}).  We put $0<\gamma<\frac12$ in
Eqs.~(\ref{massattheboundary})-(\ref{V}) and analyze the main spacetime
features.  From Eq.~(\ref{massattheboundary}) we have
$m\rightarrow 0$, i.e., $m=0$ in the limit.  Equation~(\ref{mp})
yields $m_p=\infty$, the proper mass is infinite.  From
Eq.~(\ref{lp}), the total proper length $l_p$ is then infinite. From
Eq.~(\ref{area}) the surface area is $A= 0$, and the area radius
of the boundary is $r_0=0$.  From Eq.~(\ref{V}), the proper volume for
$0<\gamma<\frac13$ is zero, $V_p=0$, for $\gamma=\frac13$ it is finite
nonzero, $V_p=\frac{32\mu^3}{3}$, in which case it is a string or a
rope with zero cross section area and infinite length but finite
volume, and for $\frac13<\gamma<\frac12$ it is infinite, $V_p=\infty$.

\begin{figure}[h]
\begin{center}
\includegraphics[scale=0.5]{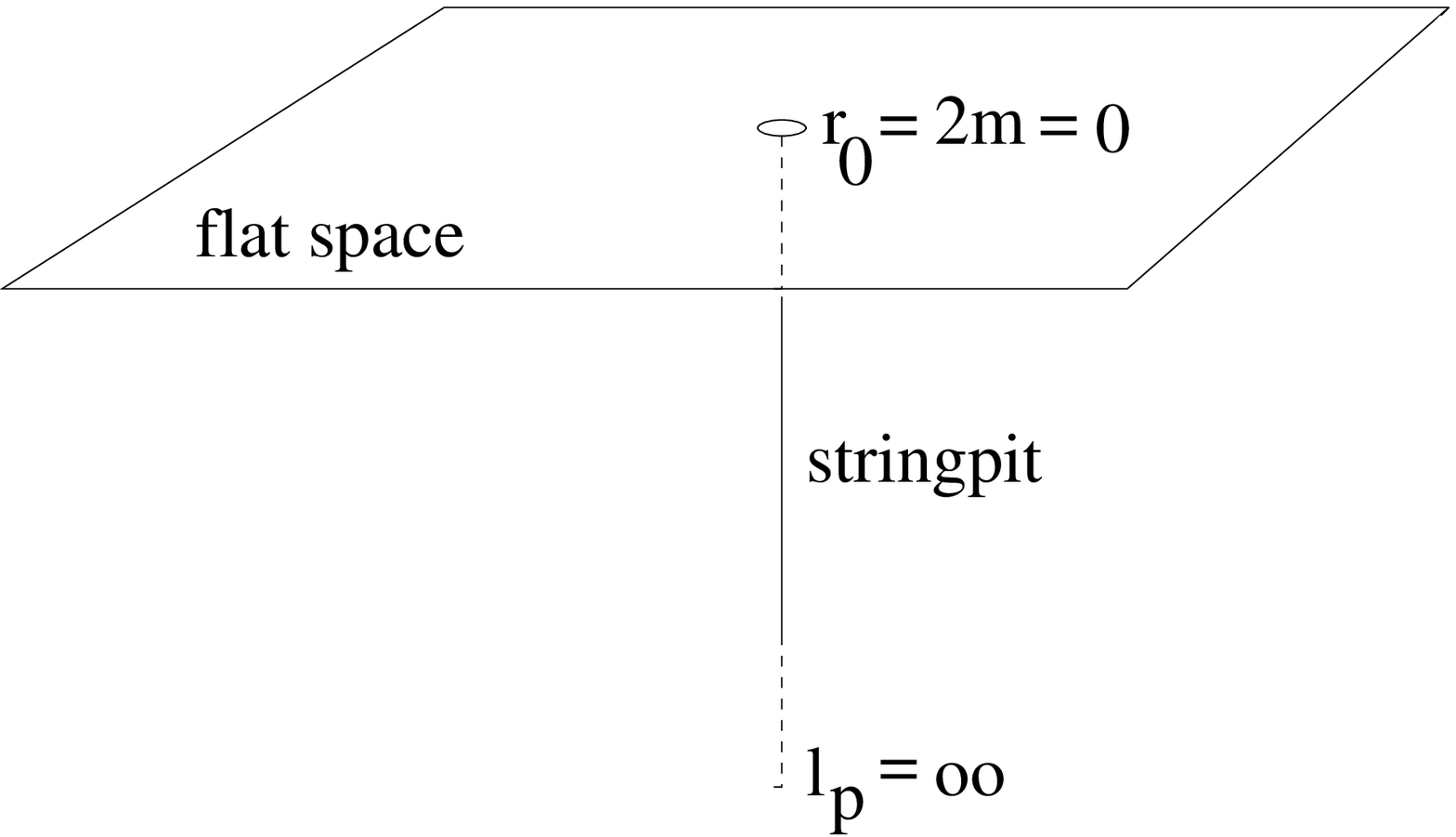}
\caption{
A $t={\rm constant}$ and $\theta=\frac\pi2$ space representation of
the spacetime given by the Zel'dovich-Letelier-Schwarzschild star with
proper mass $m_p={\rm infinite}$, actually $m_p(1-b)^{\gamma}=\mu$
with $0<\gamma<\frac12$ and $b=1$ and spacetime mass $m=0$.  This
class has $r_0=2m$, so it is a quasiblack hole, an atypical one, as it
satisfies $r_0=2m=0$.  The space inside is a a region of matter packed
at the highest level, composed of a pit made of a one-dimensional
string with infinite proper length, hung from a point with zero area
$A=0$ and $r_0=0$, and, depending on the parameter $\gamma$, with
zero, finite, or infinite volume, hanged from a point with $r_0=0$,
which opens up to a massless $m=0$ Minkowski spacetime, i.e., a flat
space.  The point $r_0=0$ yields the singular horizon of the
quasiblack hole and joins the almost detached semi-infinite string to
the rest of the space.  Note that although $m=0$ and $r_0=0$ their
ratio is finite, as $\frac{2m}{r_0}=1$.  This object has maximal mass
defect, indeed infinite mass defect.  The representation of the
semi-infinite string pit solution shows that the solution is a totally
squashed Wheeler bag of gold, although infinite in this class.
}
\label{case2}
\end{center}
\end{figure}

Thus, the full spacetime can be understood as follows.  The inside
solution is made of a one-dimensional string, with infinite length,
zero area, and zero, finite, or infinite volume depending on the
specific $\gamma$.  That the inside spacetime is a one-dimensional
string can be also seen from the conical form of the inside metric,
where for $b=1$ one has that $\bar r^2(1-b)$ tends to zero as is
the case
for the range of $\gamma$ under study, and thus
the angular part disappears,
leaving a one-dimensional space, i.e., a two-dimensional spacetime.
This packed region of matter inside,
made of a lonely boundless
semi-infinite string
in a pit almost detached from the outer
spacetime hanging from a point, is the remnant of the infinite number
of strings stemming radially from $r=0$ up to $r_0$ in a hedgehog
distribution in the original Zel'dovich-Letelier interior solution.
For
the shell that joins the inside and the outside, one deduces it is now
a point as $r_0=0$.  Then, from Eq.~(\ref{shellstresses}), since $b=1$
and $r_0=0$, the tangential stresses tend to infinity, $P\rightarrow
\infty$, and thus the point $r_0=0$ is singular, a type of singular
horizon.  For the outside, one has that the spacetime is Minkowski as
$m=0$.  Thus, in a nutshell, a Minkowski exterior space hides a
semi-infinite string pit.  For a $t={\rm constant}$ and, e.g.,
$\theta=\frac\pi2$ space representation of the spacetime, see
Fig.~\ref{case2}, where it is clear that the region  packed
with matter
is a pit with a semi-infinite string hanging in the middle of flat
space.



Note also five additional important and interesting properties
of this class of string pit solutions.
First, the interior mass of the Zel'dovich-Letelier-Schwarzschild
star \cite{z,letelier,barriolavilenkin,guend,isr} in
this limit is hidden to the outside, as it does not manifest itself
gravitationally to the outer space since $m=0$.
Second, it is also hidden because it is invisible since it is a
quasiblack hole \cite{qbh1,qbh2,qbh3,qbh4,qbh5}. Thus, it is invisible
for two reasons.  Third, although $m=0$, its ratio to $r_0$ is
finite,indeed,
$\frac{2m}{r_0}=1$, these three features
characterizing an atypical quasiblack hole.
Thus, the dynamical gravitational collapse setting in
\cite{lemos1,lemos2,lemos3,choptuik} for which
a null naked singularity, i.e., a singular horizon, forms
when $m=0$ at $r=0$ and 
$\frac{2m}{r}=1$ is also established in the static case that we are
analyzing.
Fourth, the mass defect, i.e., the proper mass minus the
energy of the assembled object given
in Eq.~(\ref{massdefect}) is $\Delta m=m_p-m=m_p=\infty$,
so we are in the presence of an object
with infinite mass defect, see also \cite{ruban,mass}. 
Fifth, it is a totally squeezed Wheeler bag of
gold \cite{wheeler,ong} if we allow
the bag to have infinite length.

The study of the geodesics in this spacetime can be done along the
lines sketched in the previous spacetime.

\subsection{A  compact stringy star at its gravitational
radius}

Here we find a  compact stringy star at its gravitational radius, with
spacetime mass $m={\rm finite}$ and the proper mass $m_p=\infty$.
This is the class $\gamma=\frac12$.

We are again interested in the limit in which $r_0\rightarrow 2m$, see
Eq.~(\ref{lr02m}), i.e., the quasiblack hole state.  It follows from
Eq.~(\ref{massattheboundary0}) that again this implies that $b\rightarrow
1$, see Eq.~(\ref{bto1}).  We put $\gamma=\frac12$ into
Eqs.~(\ref{massattheboundary})-(\ref{V}) and analyze the main spacetime
features.  From Eq.~(\ref{massattheboundary}) we have $m=\mu$,
i.e., $m$ is finite in the limit.  Equation~(\ref{mp}) yields
$m_p=\infty$, the proper mass is infinite.  From Eq.~(\ref{lp}), the
total proper length $l_p$ is then infinite. From Eq.~(\ref{area}) the
surface area $A=16\pi m^2$, which is finite, and the area radius
of the boundary $r_0=2m$ is also finite.  From Eq.~(\ref{V}), the proper
volume is infinite, $V_p=\infty$.

\begin{figure}[h]
\begin{center}
\includegraphics[scale=0.5]{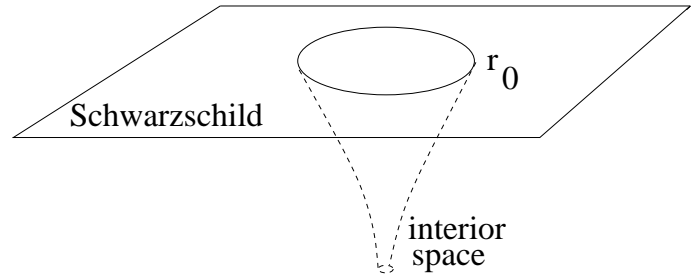}
\caption{
A $t={\rm constant}$ and $\theta=\frac\pi2$ space representation of
the spacetime given by the Zel'dovich-Letelier-Schwarzschild star with
proper mass $m_p=\infty$, actually $m_p(1-b)^{\gamma}=\mu$, with
$\gamma=\frac12$ and $b=1$, and spacetime mass $m={\rm finite}$. This
class has $r_0=2m$, so it is a quasiblack hole, a typical one, as
it satisfies $r_0=2m$ with $m$ finite.  The space inside is an
infinite volume region of matter, composed of the strings from the
original Zel'dovich-Letelier solution but now hidden in a spacetime
inside a horizon at finite nonzero $r_0$ that is singular and joins
the inside to the curved Schwarzschild exterior space. Note that
$\frac{2m}{r_0}=1$, as it should be for a quasihorizon. This object has
maximal, actually infinite, mass defect. The representation of the
compact stringy star quasiblack hole solution shows that there is no
Wheeler bag of gold in this class.
}
\label{case3}
\end{center}
\end{figure}

Thus, the full spacetime can be understood as follows. The inside
solution is made of a bulk and all the strings from the original
Zel'dovich-Letelier solution remain, but they are
now hidden in a spacetime
inside a horizon at finite nonzero area $A$ and finite $r_0$. Note
that $\bar r_0$ is infinite in this case and  the inside
space is therefore three dimensional, not one
dimensional as in the previous two
classes.  For the shell that joins the inside and the outside, one
deduces it is a sphere with radius $r_0=2m$.  Then, from
Eq.~(\ref{shellstresses}), since $b=1$, the tangential stresses tend
to infinity, $P\rightarrow \infty$, and thus the horizon $r_0$ is
a null naked horizon.  For the outside, one has that the spacetime is
Schwarzschild as $m$ is finite and not zero.  In brief, the solution
represents a  compact stringy star at the
quasiblack hole state, made of strings from $r=0$
to $r_0$, the compact star's boundary is a quasihorizon, and the
outside is Schwarzschild. This solution is not a pit.  For a $t={\rm
constant}$ and, e.g., $\theta=\frac\pi2$
space representation of the spacetime see Fig.~\ref{case3}, where it
clear that the region packed with matter has finite
boundary area radius $r_0$  and unbound
volume.

The five additional properties for this class of the compact stringy
star solution can be put in the form.  First, the interior mass of the
Zel'dovich-Letelier-Schwarzschild star
\cite{z,letelier,barriolavilenkin,guend,isr} is not hidden in this
class from the outside as it does manifest itself gravitationally to the
outer space since $m$ is finite. Second, nonetheless it is still
invisible since it is a quasiblack hole
\cite{qbh1,qbh2,qbh3,qbh4,qbh5}. Third, here $\frac{2m}{r_0}=1$ with
$m$ and $r_0$ finite, so the solution is a quasiblack hole, a typical
one in this class.  In the dynamical gravitational collapse setting
\cite{lemos1,lemos2,lemos3,choptuik} there are also cases for which
$\frac{2m}{r}=1$, with $m$ finite, characterizing the formation of a
typical black hole, not a naked singularity.  Fourth, the mass defect,
i.e., the proper mass minus the energy of the assembled object given
in Eq.~(\ref{massdefect}) is $\Delta m=m_p-m=m_p=\infty$, so we are in
the presence of an object with infinite mass defect, see also
\cite{ruban,mass}.  Fifth, this class does not resemble a Wheeler bag
of gold \cite{wheeler,ong} at all.

The study of the geodesics in this spacetime can be done along the
lines sketched in the first spacetime.


\section{Conclusions: Synopsis, thermodynamics, and connection
to other works}
\label{c}

\subsection{Synopsis of the three solutions}
\label{synopis}

The main results of this work are the finding
of two classes of 
string pit solutions with unusual interesting structures and unusual
interesting general relativistic gravitational fields. There is also
another class, a compact stringy  star solution that has standard
properties. These three classes, although obtained from an appropriate
limit of the Zel'dovich-Letelier-Schwarzschild star, stand on their
own as separate general relativistic solutions, if one wishes to
envisage them as such.  In Table~I, a summary of the main features of
the three classes parametrized by the exponent $\gamma$ is
displayed.
\vskip 0.5cm
\begin{center}
\begin{tabular}{|l|l|l|l|l|l|}
\hline
Class & ${\rm String\; pit}_1$ & ${\rm String\; pit}_2$
&  Compact stringy star \\ \hline
$\gamma$ & $0$ & $(0,\frac12)$ & $\frac{1}{2}$
\\ \hline
$m$ & $0$ & $0$  & Finite \\ \hline
$m_{p}$ & Finite & Infinite &  Infinite \\ \hline
$l_{p}$ & Finite & Infinite & Infinite \\ \hline
$A$ & $0$ & $0$  & Finite \\ \hline
$V_{p}$ & $0$ & 0, Finite, Infinite & Infinite \\ \hline
\end{tabular}
\end{center}
\vskip 0.05cm
TABLE I: {\small
The main physical features along with its values of the three
different classes of solutions, i.e., the first string pit class, the
second string pit class, and the compact stringy star class,
distinguished by the values of $\gamma$, namely, $\gamma =0$,
$0<\gamma <\frac{1}{2}$, and $\gamma =\frac{1}{2}$, are displayed. The
physical features are the spacetime mass $m$, the interior proper mass
$m_p$, the interior proper length $l_p$, the surface area at the
junction $A$, and the interior proper volume $V_p$.  There is also the
mass defect, i.e., the proper mass minus the spacetime mass or energy
of the assembled object, $\Delta m=m_p-m$, which can be taken directly
from the displayed values.
}
\vskip 0.5cm

Some properties of the solutions in the three classes found here are
as follows.
(1) For the two first classes, comprising the string pit solutions that
arise as the quasiblack hole limit of the
Zel'dovich-Letelier-Schwarzschild star, in spite of having in the core
a nonzero mass $m_p$, which in one class is arbitrarily large, this
mass is hidden, as it does not manifest itself gravitationally to the
outer spacetime since $m=0$. The third class, the compact stringy star
solution, does not possess this property, the outer spacetime is
Schwarzschild, and it has a finite nonzero $m$.
(2) The three classes of solutions are invisible to the exterior since
they are quasiblack holes, and as such no particle or light emanates
from them.
(3) The three class of solutions obey the quasiblack hole condition
$\frac{2m}{r_0}=1$. The two classes of string pit solutions are
remarkable because they not only obey this condition, but have in
addition $m=0$ and $r_0=0$, and they are indeed atypical quasiblack
holes.  These are the static solutions akin to the naked
singularities, i.e., singular horizons, that form in dynamical
gravitational collapse when $m=0$ at $r=0$, and $\frac{2m}{r}=1$. The
class of the compact stringy star solution has finite $m$ and finite
$r_0$ and represents typical quasiblack holes, akin to the black holes
that form in dynamical gravitational collapse with $m$ finite and some
finite horizon radius $r$. Surprisingly, these classes of static
solutions yield the same spectrum that appear in critical
gravitational collapse.  Indeed, in gravitational collapse there are
solutions that yield naked null singularities that correspond here to
the two string pit classes of solutions that also have naked null
singularities, there are solutions that yield black holes that
correspond here to the class of compact stringy stars at the
quasiblack hole limit, and the solutions that disperse away in
gravitational collapse here are the static
Zel'dovich-Letelier-Schwarzschild stars that we considered initially.
(4) In the three classes, the mass defect, i.e., the proper mass $m_p$
minus the spacetime mass $m$ of the assembled objects, is maximal. In
the first class of string pit solutions the mass defect is finite and
maximal, and in the other two classes, i.e., the second class of
string pit solutions and the compact stringy star solution, it is a
superstrong mass defect, it is infinite.  Moreover, as far as the
gravitational mass defect is concerned, we have obtained general
results, indeed we have shown that the maximal mass defect result can
be obtained without specifying any equation of state. The
Zel'dovich-Letelier-Schwarzschild star is a realization of the general
result.
(5) For the two first classes, i.e., the two string pit solutions, one
finds Wheeler bags of gold, albeit totally squashed ones. The third class,
the compact stringy star solution, has no bag.
The finding of these three classes and the intepretation of them has
benefited from several works
\cite{z,letelier,barriolavilenkin,guend,isr,qbh1,qbh2,qbh3,
qbh4,qbh5,lemos1,lemos2,lemos3,choptuik,ruban,mass,wheeler,ong}.

\subsection{Thermodynamics of the three solutions}
\label{th}

It is also of interest to study the thermodynamic behavior of the
string pits and stringy compact star quasiblack hole solutions which
have an equation of state $p_{r}=-\rho$ for their interior.  The
appropriate thermodynamic formalism has been developed and is ready
\cite{ent,ext}.  When studying the thermodynamics of each system we
suppose that a local temperature $T$ has been assigned to it.  We deal
with the main features of the temperature distribution and the entropy
of the solutions, one at a time.

In relation to the temperature, in a gravitational system one has the
Tolman temperature formula $T=\frac{T_0}{\sqrt{1-\frac{2m(r)}{r}}}$,
for some local temperature $T$ of the system, which in general is
different for each sphere with radius $r$, i.e., $T=T(r)$, and a
temperature at infinity $T_0$, say, which has some constant value.
Now, throughout the interior, for the string pit and the stringy star
solutions one has that $\frac{2m(r)}{r}$ is a constant, indeed,
$\frac{2m(r)}{r}=b$, and since $T_0$ is a constant, the whole interior
solution, for any $r$, has the same local temperature $T(r)=
\frac{T_0}{\sqrt{1-b}}$, a constant, so we can speak of an
isothermal interior.  Moreover, considering $T$ of the system finite.
in the limit $b=1$ we then have $T_0=T\sqrt{1-b}=0$, so a remote
observer will measure a vanishing temperature $T_0$, $T_0=0$, for
these solutions.  Usually, zero $T_0$ is a feature of an extreme
quasiblack hole or of an extreme black hole.  However, the systems we
have analyzed are nonextremal. Thus, it seems that the string pits and
the sringy star combine features of nonextremal and extremal
quasiblack holes and black holes \cite{ent,ext}.

In relation to the entropy, the entropy of a nonextremal quasiblack
hole is $S=\frac 14 A_+$, where $A_+=4\pi r_+^2$ is the horizon area,
with $r_+$ being the horizon radius. Thus, it is the Bekenstein-Hawking
formula for the entropy.  Taking the limit $b=1$,  $r_0$ is the
radius of the system at the quasihorizon, $r_0=r_+$.  For the string
pits when $b=1$, $r_0=r_+=0$, and thus $S=0$. Thus, entropically speaking,
it has an extremal quasiblack hole behavior \cite{ext}.  For the
stringy star, $r_0=r_+$ is finite and thus $S=\frac 14 A_+$,
which is a typical
nonextremal quasiblack hole behavior.

\subsection{Connections of the three solutions to other works}
\label{f}

There are relevant and related solutions to the Zel'dovich-Letelier
interior that have the mass function $m(r)$ obeying
$\frac{2m(r)}{r}=b$ but are not string dust, i.e., the equation of
state is not $p_r=-\rho$. These interior related solutions when
matching to a Schwarzschild exterior do not give the
Zel'dovich-Letelier-Schwarzschild stars that we have been treating.

One example of such solutions is notable. In general relativity, using
the Toman-Oppenheimer-Volkoff equation abbreviated usually to TOV
equation, for the hydrostatic equilibrium of a spherically symmetric
configuration with matter having an equation of state of the form
$p=-q\rho$, with $p$ a perfect fluid pressure and $q$ some number, one
finds that the energy density $\rho$ is proportional to $\frac1{r^2}$
and $m(r)$ indeed obeys $\frac{2m(r)}{r}=b$, as reported with
distinction in \cite{klein,miszapo,chandra} and wrapped up and
developed in \cite{chavanis,bg}.  This general relativistic solution
is called an isothermal perfect fluid solution because it is a
generalization, albeit a nonisothermal one, of the Emden equation for
isothermal spheres made of an ideal gas in Newtonian gravitation.

There are other instances where $\frac{2m(r)}{r}=b$ appears.  We
mention regular black holes \cite{dymnikova1992} where the solutions
besides the $p_r=-\rho$ equation also have tangential pressure $p_t$
support, and general relativistic solutions coupled to nonlinear
electrodynamics which have similar features \cite{nl}. We also allude
to a thermodynamic treatment of the semiclassical degrees of freedom
of a black hole which yields the expression $\frac{2m(r)}{r}=b$ as a
plausible equation \cite{berezin}.

One can only express wonder at all these interconnections from so many
different settings.


\begin{acknowledgments}
JPSL thanks Funda\c c\~ao para a Ci\^encia e Tecnologia - FCT,
Portugal, for financial support through Grant~No.~UIDB/00099/2020.
OBZ thanks Kazan Federal University for a state grant for scientific
activities.
\end{acknowledgments}

\appendix

\section{Geodesics in the spacetime of case A. A finite string in a
pit, i.e., a string pit, almost detached from spacetime hanging from a
point}
\label{geo}

\subsection{Timelike geodesics}

Here, we study radial timelike geodesics of the spacetime
of case A. above, i.e., a
finite string in a
pit almost detached from spacetime hanging from a point,
with $m=0$ and $m_p={\rm finite}$.
A radial geodesic has as one of its equations the 
equation
$\left(1-\frac{2m(r)}{r}\right)\dot t=E$, where the dot indicates
derivative with respect to proper time $\tau$ and
$E$ is a constant
representing
the energy per unit mass of the massive test particle
along the geodesic. The other equation is
$\left(1-\frac{2m(r)}{r}\right)\dot t^2
-\frac{\dot r^2}{1-\frac{2m(r)}{r}}=1$.
Combining the two, one gets
$\dot r^2=E^2-1+\frac{2m(r)}{r}$.
Thus, $d\tau=\pm\frac{dr}{\sqrt{ E^2-1+\frac{2m(r)}{r}}}$.
For the inside $\frac{2m(r)}{r}=b$, so letting 
a massive test particle fall from some $r$
in the inside region
to the center gives the proper time 
$\tau =\int_{0}^{r}
\frac{dr^\prime}{\sqrt{ E^2-1+b}}$, which yields
$\tau =
\frac{r}{\sqrt{ E^2-1+b}}$.
If the  particle comes from $r_0$
with $E\geq 1$,
then
\begin{equation}
\tau_{\rm in} =
\frac{r_0}{\sqrt{ E^2-1+b}}\,.
\end{equation}
In the limiting spacetime, $b\to1$ and $r_0\to0$ yields
$\tau_{\rm in} =0$.

\subsection{Null geodesics}

Here, we study radial null geodesics of the spacetime
of case A. above, i.e., a
finite string in a
pit almost detached from spacetime hanging from a point,
with $m=0$ and $m_p={\rm finite}$.
For a null geodesic $ds^2 =0$.
Thus, the time between the center and some $r$ 
is
$t=\int_{0}^{r}\frac{dr^\prime}{1-\frac{2m(r^\prime)}{r^\prime}}$.
For the inside $\frac{2m(r)}{r}=b$,
so the time between the center and the boundary $r_0$
is
\begin{equation}
t_{\rm in}=\frac{r_0}{1-b}\text{.}
\end{equation}
In the limiting spacetime, $b\rightarrow1$ and $r_{0}\rightarrow0$,
and 
taking into account Eqs.~(\ref{mrzl}) and~(\ref{mprzl})
with $m_p$ finite, one has
$t_{\rm in}\rightarrow \infty$ with $\frac1{\sqrt{1-b}}$.
The time between $r_0$ and some $r_1>r_0$
is 
$t_{\rm out}
=
\int_{r_0}^{r_1}\frac{dr}{1-\frac{
2m(r_0)}{r}}=\int_{r_0}^{r_1}\frac{dr}{1-
\frac{br_0}{r}}
=
\int_{r_0}^{r_1}\frac{drr}{r-br_0}=r_1-
r_0+br_0\ln \frac{
r_1-br_0}{r_0(1-b)}$, i.e., 
\begin{equation}
t_{\rm out}
=r_1-
r_0+br_0\ln \frac{
r_1-br_0}{r_0-br_0}
\end{equation}
When $b\rightarrow 1$, $t_{\rm out}\rightarrow \infty$.
The divergences of $t_{\rm in}$
are stronger than those of $t_{\rm out}$.

\subsection{Redshift and blueshift of light}

Now we analyze the redshift and blueshift of light
of the spacetime
of case A. above, i.e., a
finite string in a
pit almost detached from spacetime hanging from a point,
with $m=0$ and $m_p={\rm finite}$.
According to the standard formulas,
\begin{equation}
\omega =
\frac{\omega_{0}}
{\sqrt{1-\frac{2m(r)}{r}}}
\,,
\end{equation}
where $\omega $ is the frequency measured by a local static observer
at a given point $r$
and
$\omega_{0}$ is a constant. Inside, one has
$1-\frac{2m(r)}{r}=1-b$, and outside
$1-\frac{2m(r)}{r}
=1-\frac{2m}{r}$.
Thus, inside, for $r\leq r_0$, the frequency
of light does not change as it propagates.
On the other outside, $r\geq
r_0$,
one can put $\omega _{0}\equiv \omega _{\infty }$,
the frequency at infinity. 
Therefore, when light goes from $r_0$ to an even 
larger $r$,
it will arrive 
at infinity with $\omega _{\infty }=\omega \sqrt{1-b}$.
In the
limit $ b\rightarrow 1$, we have an infinite
redshift, as expected for a quasiblack hole.
If light with finite $\omega _{\infty }$ comes from infinity and
enters the inner region, it has
$\omega =\frac{\omega_\infty}{\sqrt{1-b}}$ there.


\end{document}